\documentclass[
]{ceurart}

\usepackage{listings}
\newcommand\AND{{\scshape and}\xspace}
\newcommand\NAND{{\scshape nand}\xspace}

\newcommand\NOT{{\scshape not}\xspace}

\usepackage{graphicx}
\usepackage{listings}
\usepackage{mathtools}
\usepackage{wrapfig}

\begin{document}

%%
%% Rights management information.
%% CC-BY is default license.
\copyrightyear{2023}
\copyrightclause{Copyright for this paper by its authors.
  Use permitted under Creative Commons License Attribution 4.0
  International (CC BY 4.0).}

%%
%% This command is for the conference information
\conference{ESWC '23: Trusting Decentralised Knowledge Graphs and Web Data Workshop – TrusDeKW, May 28--June 1, 2023, Hersonissos, Greece}

%%
%% The "title" command
\title{RDF Surfaces: Computer Says No}

\tnotemark[1]

%%
%% The "author" command and its associated commands are used to define
%% the authors and their affiliations.
\author[1,2]{Patrick Hochstenbach}[%
orcid=0000-0001-8390-6171,
email=Patrick.Hochstenbach@UGent.be,
url=https://patrickhochstenbach.net/
]
\cormark[1]

\address[1]{Ghent University Library, Rozier 9, Ghent, 9000, Belgium}
\address[2]{Ghent University IDLab - ELIS, Technologiepark-Zwijnaarde 122, Zwijnaarde, 9052, Belgium}

\author[2]{Jos De Roo}[%
orcid=0000-0001-8862-0666,
email=Jos.DeRoo@UGent.be
]

\author[2]{Ruben Verborgh}[%
orcid=0000-0002-8596-222X,
email=Ruben.Verborgh@UGent.be,
url=https://ruben.verborgh.org/
]

%% Footnotes
\cortext[1]{Corresponding author.}

%%
%% The abstract is a short summary of the work to be presented in the
%% article.
\begin{abstract}
Logic can define how agents are provided or denied access to resources, how to interlink resources using mining processes and provide users with choices for possible next steps in a workflow. These decisions are for the most part hidden, internal to machines processing data. In order to exchange this internal logic a portable Web logic is required which the Semantic Web could provide. Combining logic and data provides insights into the reasoning process and creates a new level of trust on the Semantic Web. Current Web logics carries only a fragment of first-order logic (FOL) to keep exchange languages decidable or easily processable. But, this is at a cost: the portability of logic. Machines require implicit agreements to know which fragment of logic is being exchanged and need a strategy for how to cope with the different fragments. These choices could obscure insights into the reasoning process. We created RDF Surfaces in order to express the full expressivity of FOL including saying explicitly `no'. This vision paper provides basic principles and compares existing work. Even though support for FOL is semi-decidable, we argue these problems are surmountable. RDF Surfaces span many use cases, including describing misuse of information, adding explainability and trust to reasoning, and providing scope for reasoning over streams of data and queries. RDF Surfaces provide the direct translation of FOL for the Semantic Web. We hope this vision paper attracts new implementers and opens the discussion to its formal specification.
\end{abstract}

%%
%% Keywords. The author(s) should pick words that accurately describe
%% the work being presented. Separate the keywords with commas.

%%
%% This command processes the author and affiliation and title
%% information and builds the first part of the formatted document.
\maketitle

\section{Introduction}
\label{Introduction}

RDF \cite{RDF} is the standard for modeling data on the Semantic Web. From a syntactic viewpoint, RDF is a simple data model for expressing relations between triples where each triple is a first-class web object. From a semantic viewpoint, a triple is an assertion expressing what is believed to be \emph{true}. Combinations of triples form a logical conjunction (\AND). Any combination of resources on the Semantic Web creates a universal \AND statement \cite{RDFSemantics}. This is not unusual as the majority of database systems assert truth similarly.

Human and software agents interpret data and use internal logic to turn these assertions into decisions regarding what data to \emph{trust} or not (negation), provide \emph{options} to select data to ingest or compute (disjunction) and make \emph{inferences} from this data (implication). Current Web logics in the form of rule and ontology languages provides insights into the reasoning process but only carries a fragment of first-order logic (FOL). These fragments of FOL are tuned to make the processing of Web logic decidable or easily processable, at the cost of portability.
\emph{Portability} is the ability to represent data and logic independently irrespective of the processing environment. RDF is portable, any RDF triple expresses the same data and meaning irrespective of the source. Adding Web logic, this becomes much harder: machines must first agree on an entailment regime before they know what each triple represents. In this position paper, we present our case for a portable Web logic language extending RDF semantics called \emph{RDF Surfaces}, which is as powerful as FOL. RDF Surfaces is able to group RDF data within surfaces and provide semantics to say `no': expressing an explicit scope and classical negation.

\section{Background}
\label{background}

We started our research into portable logic in relation to our involvement in the SolidLab Flanders \cite{SolidLab} and Mellon-funded project ``Scholarly Communications in the Decentralized Web'' \cite{Mellon}. In both projects, a decentralized network of \emph{data pods} exist on which individuals store (personal) data. In the case of SolidLab, data are documents Flemish citizens share with the government or businesses. In the case of the Mellon project, data are research artifacts such as scholarly articles and research data  that are shared within a research community. Both projects are investigating how automated systems can assist in managing these types of data, interlink data with existing (external) resources, and provide enforcement of data policies using logic-based rules. In each use case, automated systems can't assume that only one actor creates and manages these rules. In a decentralized world, many agents can set their own requirements of what should or shouldn't happen with data. For instance, in the case of data policies, different actors on a personal, institutional, and governmental level can define the permissions and prohibitions (re)using data. These policies can overlap and possibly contradict themselves. Such clashes can benefit from being detected at an early stage and not at run time. Another challenge is the single language problem. Machines should be capable to implement potentially heterogeneous collection policy languages with many possible dialects. The strength of RDF is that the data model for these variations is portable, there is a common format that can express the overlap between the different policy languages. But, it is not the case that the logic -- what all RDF triples entail -- is portable. Without knowing which kind of inferences are possible (using entailment regimes such as RDFS, OWL-DL, OWL2, Notation3,..), it is possible that designers of policies can reach other conclusions than the consumer (the policy enforcer on the pod). For these use cases, we need a common portable logic, with a common syntax and semantics.

\section{RDF Surfaces}
\label{RDF Surfaces}

Only two extensions are required to the semantics of RDF to make it expressable as FOL: a notion of a (possible nested) \emph{surface} to group zero or more RDF graphs with a truth-functional type, and a notion of \emph{graffiti} as marks on such as a surface with the function of existentially quantified variables. 

\textbf{Surfaces} can be regarded as nestable virtual sheets of paper on which RDF graphs are written. A \emph{positive surface} asserts all triples written on it; as with RDF, multiple triples form a logical conjunction~(\AND). The default surface is positive. All existing RDF graphs are regarded to be on the default positive surface. A \emph{negative surface} negates all triples written on it, and multiple triples form a negation~(\NOT) of a conjunction~(\AND).\footnote{Individual triples are not necessarily negated; only their conjunctions.} With logical connectives \AND and \NOT, any truth-functional statement can be created by composition, similar to how logical gates on a computer chip can be created by combining \NAND gates.

\textbf{Graffiti} are marks on a surface representing quantified logical variables. Graffiti marks on a positive surface are interpreted as \emph{existentially} quantified variables, this is how blank nodes are currently interpreted in RDF. The difference between blank nodes and graffiti marks is that blank nodes act as \emph{coreferences} to graffiti marks, and that every surface contains its own unique set of graffiti marks. Transporting marks to a new surface requires creating new graffiti marks (not relabeling or merging graffiti marks, as is the case with blank nodes). Using De~Morgan's duality\footnote{%
  $\lnot \exists x\colon P(X)  \equiv \forall x\colon \lnot P(x)$
  }, graffiti marks on a~negative surface are interpreted as \emph{universally} quantified variables.

Given the notion of a surface and graffiti, we define an \textbf{H-graph} as the combination of a (typed) surface $S$, graffiti $Gr$, and a graph $H$ which is again an H-graph. Each RDF graph is an H-graph on the default positive typed surface. The empty H-graph is regarded as a tautology on a positive-typed surface and a contradiction on a negative-typed surface. By combining and nesting H-graphs any truth-functional statement can be created. H-graphs have similar semantics as the alpha and beta Existential graphs of Pierce\footnote{\url{https://plato.stanford.edu/entries/peirce-logic/\#AlphSyst}} which are as expressive as FOL\cite{Zeman}\cite{Roberts}. Our position is that these two added primitives (surfaces and graffiti marks) are a much more economical way to express FOL than in OWL Full, plus they are an immediate natural extension to RDF semantics.

We need a notation to transport the H-graphs over the Web and require an RDF syntax for that. This syntax should provide a way to scope RDF graphs and codify the graffiti marks. Our first choice was to use Notation3 \cite{Notation3}, because it provides syntactic support for nested quoted graphs (surfaces), built-ins (to reason with surface types), and lists (codifying the graffiti marks). Using Notation3, an H-graph is expressed using graffiti marks as the subject list, a typed surface as a predicate, and an H-graph as the object. In our implementations, the surface predicate is implemented as a built-in to allow for reasoning with H-graphs.

Listing \ref{list:n3ex} provides an example of an H-graph with semantics
 $
  \forall s\colon \mathit{learns}(s,Physics)
    \Rightarrow
      \mathit{reads}(s,Newton)
      \lor
      \mathit{reads}(s,Einstein)
  $
by stating it as
  $
    \forall s\colon \lnot(
      \mathit{learns}(s,Physics)
      \land
      \lnot\mathit{reads}(s,Newton)
      \land
      \lnot\mathit{reads}(s,Einstein)
    )
  $
which means ``Anyone that learns physics reads Newton or Einstein (or both)''.

\begin{lstlisting}[float,label=list:n3ex,caption=The RDF Surfaces semantics of a material implication with a disjunction using the Notation3 syntax.]
(_:S) log:onNegativeSurface {
    _:S :learns :Physics .
    () log:onNegativeSurface { _:S :reads :Newton } .
    () log:onNegativeSurface { _:S :reads :Einstein } .
} .
\end{lstlisting}

For a detailed introduction to RDF Surfaces we refer to our RDF Surface Primer \cite{Primer}.

\section{Related Work}
\label{RelatedWork}

The case for portable Web logic is not new.  Already in the 2000s Berners-Lee \cite{Swell} called for the development of a language on top of RDF in his Semantic Web Logic Language (SWeLL) proposal. SWeLL was imagined as a unifying language acting as a logical bus to ``allow any web software to read and manipulate data published by any other web software''. For all logical relations to be expressed, SWeLL had to extend RDF by including negation and explicit quantification.

A similar portability argument was given in 2009 by Hayes in his ISWC invited talk \cite{Blogic}. According to his argumentation, the idea of Web logic portability has not been achieved due to the layering of logic in the Semantic Web stack. Different layers of the stack cannot guess which entailment regimes are used when receiving data. Without this information, two independent machines cannot arrive at the same conclusions given a model. As a first step towards a solution for this conundrum, Hayes presented RDF Redux which provides a syntax for expressing Web logic with FOL semantics. His ideas were very influential for our work and led to our current research into RDF Surfaces.

There are two main reasons why classical negation has thus far been avoided. The first reason is negation opens the door to create \emph{paradoxes} on the Semantic Web. One of the main motivations of Berners-Lee et al. for N3Logic \cite{N3Logic} was to introduce the ability to compare Web documents and make inferences about assertions in each document. For this reason, a quoting mechanism was introduced. Quoting in combination with classical negation can lead to paradoxes when assertions in resources are self-referential. This self-referential problem resembles the liar paradox in the philosophy of language: ``This sentence is false''. To avoid paradoxes, Scoped Negation As Failure (SNAF) was introduced to N3Logic and acts as a monotonic version of Negation As Failure (NAF). SNAF disallows self-referential negative statements about resources and defines a scope for negation by the absence of information (Scope + NAF = SNAF). However, SNAF cannot avoid semantic inconsistencies nor has a syntactic mechanism for expressing them. With RDF Surfaces we introduce a formalism requiring negative facts to be explicitly expressed as negative surfaces, without using (S)NAF. Inconsistencies can be expressed as an assertion plus a negative surface thereof. These inconsistencies still need to be detected which leads to our second reason.

Classical negation is also avoided due to combining the expressive power of logics with negation and \emph{decidability} problems. \emph{Satisfiability} is a decidability problem to find inconsistencies. \emph{Completeness} is a decidability problem, related to satisfiability, to find all valid statements from a knowledge base. \emph{Halting} is a decidability problem requiring a machine to stop in a finite time.  Logics containing logical variables and negation can be as powerful as first-order logic (FOL) for which it is proven \cite{Trakhtenbrot} solving any of these problems is undecidable. Three options are available for dealing with undecidability for which, alas, only two can be chosen: (P) portable, allow any ontology input, (C) be complete, (H) always halt \cite{Hogan}. OWL 2 Full is an example of a Web logic that can express any ontology but is undecidable. OWL 2 Description Logics drops option (P) and creates decidable fragments of FOL and reasoners can always return all answers and always halt, i.e. (C+H). This choice raises the question of what fragment of FOL to choose. This is at the cost of Web portability. There are at least as many choices as there are OWL Description Logics. Reasoners that cannot guess which fragment to choose to have to implement them all to provide portability. Reasoners should be as expressive as FOL in order to do so. In general, if we want portable Web logic, option (P) seems to be unavoidable. We argue: any of the (P+C), (P+H), and (C+H) axis are valuable depending on the use case. A reasoner as powerful as (P+C) or (P+H) is required to implement a portable Web logic, but the logic could be 'tunable' for specific use cases. A `tuned down' (C+H) reasoner could implement OWL Description Logics as syntactic sugar. (P+C) and (P+H) can provide a portable logic for reasoning over decentralized sources. Undecidability is a hard problem, but unavoidable in the latter case. The logic provided by RDF Surfaces, we argue, can be created with fewer primitives than OWL 2 Full and is in line with a large body of research on FOL. Undecidability itself does not stop the Web from evolving as can be seen from the popularity of Web programming languages such as JavaScript.

\section{Conclusion and future work}
\label{Conclusion}

We believe that RDF Surfaces provides an expressivity comparable to Peirce's alpha and beta graphs and FOL. We still need proof that a complete mapping from RDF Surfaces to FOL (or Peirce's graphs) exists. This is part of our current research. Existential rules have shown already to be sufficient to implement the inference rules of RDFS, OWL-RL, and OWL-EL \cite{Tomaszuk,Arndt,Kroetzsch}. These existential rules were implemented in the Notation3 language, but we are porting them to RDF Surfaces. We started to port the Notation3 language itself in RDF Surfaces as ongoing work.\footnote{\url{https://github.com/eyereasoner/Notation3-By-Example/tree/main/examples/n3s}}

The expressivity of RDF Surfaces goes beyond what is provided by OWL Description Logics. It could be impractical to express every inference rule solely in the language of first-order logic. The Notation3 language provides built-in functions and relations for arbitrary operations on RDF graphs of practical assistance to the programmer. A pure subset of these functions (deterministic and without side effects) could form the basis for an assembly language of the web. In addition to Web portability, use cases for RDF Surfaces can be found in implementing policy languages where one would explicitly want to express what is regarded as misuse of information and reason over that. RDF Surfaces can provide the semantics to limit the scope of reasoning over RDF data streams or limit the scope for queries and provide alternatives to query whatever is in reach \cite{Hartig}.

The authors are currently experimenting with reasoners implementing RDF Surfaces along the (P+H) \cite{EYE} and (C+H) \cite{Latar} axis and formed a W3C Community Group \footnote{\url{https://www.w3.org/community/rdfsurfaces/}} to define the semantics and find new implementers.

%%
%% The acknowledgments section is defined using the "acknowledgments" environment
%% (and NOT an unnumbered section). This ensures the proper
%% identification of the section in the article metadata, and the
%% consistent spelling of the heading.
\begin{acknowledgments}
This work is funded by the Andrew W. Mellon Foundation (grant number: 1903-06675) and supported by SolidLab Vlaanderen (Flemish Government, EWI, and RRF project VV023/10). The authors thank Dörthe Arndt and Ruben De Decker for several insightful discussions about logic and applications of RDF Surfaces. 
\end{acknowledgments}

%%
%% Define the bibliography file to be used
\bibliography{paper}

\end{document}